\documentclass[11pt]{amsart}
\usepackage{geometry}
\usepackage{graphicx}
\usepackage{amssymb}
\usepackage{epstopdf}
\title{Quantum-field theories as representations of a single $^\ast$-algebra}

\author{A. Raab\\ 
Hyltans Sj\" ov\" ag 12, 43539 M\" olnlycke, Sweden,\\
andreas.raab.mail@web.de}

\date{\today}

\begin{document}

\newcommand*\C{\mathrm{l}\hspace{-2mm}\mathrm{C}}
\newcommand*\R{\mathrm{I}\!\mathrm{R}}
\newcommand*\N{\mathrm{I}\!\mathrm{N}}
\newcommand*\WF{\mathrm{WF}}

\begin{abstract}
We show that many well-known quantum field theories emerge as representations of a single $^\ast$-algebra. These include free quantum field theories in flat and curved 
space-times, lattice quantum field theories, Wightman quantum field theories, and string theories. We prove that such theories can be approximated on
lattices, and we give a rigorous definition of the continuum limit of lattice quantum field theories.
\end{abstract}

\maketitle

\section{Introduction}
\label{INTRO}

The Wightman distributions play a fundamental role in Wightman quantum field theories (Wightman QFTs) \cite{str00}. The reconstruction
theorem demonstrates that knowledge of the Wightman distributions is sufficient to obtain a unique Wightman QFT. In
particular, the Wightman distributions define a state of a Borchers-Uhlmann (BU) algebra, and the
associated Wightman QFT is obtained as a representation of a BU-algebra from that state \cite{bor62,uhl62}. 
Interestingly, interacting as well as non-interacting Wightman QFTs emerge as representations of the same
BU algebra. 

Although the Wightman axioms provide a remarkably successful framework in Minkowski space-time, they cannot be generalized
to curved space-times in a straightforward manner. This is one of the motivations to study QFT in an algebraic framework, and substantial progress has been achieved 
with this approach in recent years. An overview of the field can be found in Refs. \cite{hol10,bru03}, and we will discuss more details
of those achievements in Sec. 5. 
However, the approach taken in Ref.  \cite{hol10} yields an interesting aspect that results from the generalization of the axiomatic approach to curved space-times.
Starting point of the axiomatic approach in Ref.  \cite{hol10} is a free $^\ast$-algebra, $\mbox{Free}(M)$, of quantum fields 
on a background structure, $M$, which, amongst others, refers to
a globally hyperbolic space-time. The quantum-field algebra, 
$A(M)$, is obtained by factoring $\mbox{Free}(M)$ by a set
of relations, in which coefficients of an operator-product
expansion (OPE) play a fundamental role, i.e. there exists a
$^\ast$-homomorphism, $\pi: \mbox{Free}(M) \to A(M)$, which 
essentially is defined by properties of the OPE coefficients. 
The set of states of the theory, $S(M)$, is
further constrained to support the OPE, positivity, and a microlocal spectrum condition. However, if we construct a representation of
$A(M)$ from a state $\omega \in S(M)$, then we basically obtain a
representation of $\mbox{Free}(M)$ with respect to the state
$\omega \circ \pi$. The set of states of the QFT can therefore 
also be seen as a subset of the set of states over
$\mbox{Free}(M)$, which is obtained by appropriate constraints.
Quite different QFTs can therefore emerge as
representations of the same $^\ast$-algebra, $\mbox{Free}(M)$, and we note that $\mbox{Free}(M)$ plays an analogous role
as the BU-algebras in Wightman QFTs. 
We pick up this idea and explore it in a more general approach in this paper.

In Sec. 2.1, we introduce the general setting, in which we define the terms test-function space, Q-map, and Q-theory. Q-maps are
mathematical generalizations of quantum fields that lack specific physical properties, and Q-theories are corresponding generalizations of QFTs. 
We use Q-maps and Q-theories as purely technical devices to construct appropriate $^\ast$-algebras, from which QFTs arise as representations.
In particular, as it is the case for quantum fields in Wightman QFTs, for example, the image of a Q-map is a set of operators in a $^\ast$-algebra,
and a Q-theory is the continuous representation of the polynomial algebra generated by the operators in the image of a Q-map.
In Sec. 2.2 we further define in which sense a Q-theory emerges from a representation of a $^\ast$-algebra 
(i.e. the polynomial algebra generated by a Q-map),
and in Sec. 2.3 we introduce a Q-map whose polynomial algebra, $\mathcal{A}_0$,  can be represented in any Q-theory, so that the Q-theory 
emerges from that representation. We also prove that Q-theories can be approximated on lattices, and we give
a rigorous definition of the continuum limit. In the remaining part of the paper we
show that many well-known QFTs are Q-theories that emerge as representations of $\mathcal{A}_0$.
These are the central results of this paper.

We discuss in Sec. 3 the applicability of the approach to hermitian scalar Wightman QFTs, and in Sec. 4 we discuss free
scalar QFTs and Dirac QFTs in curved space-times. In Sec. 5, we continue with the discussion of perturbatively interacting quantum
fields in curved space-times, and we also relate to recent developments in algebraic QFT in more detail, c.f. Refs. \cite{hol10,bru03}.
In Sec. 6, we consider string theories and lattice QFTs, and we discuss the continuum limit of
lattice QFTs more concretely. 

\section{Q-maps and Q-theories}
\label{GENERAL}

\subsection{General setting}
Let us begin with the definition of some general terms.  
In Wightman QFTs, quantum fields are defined as operator-valued distributions over a test-function space.  We choose a similar approach:
\begin{enumerate}
\item  A conjugation, $C$, on a complex vector space, $V$,
is an antilinear map satisfying
$C^2 = 1$ and $C( av + bw) = \bar{a} C(v) + \bar{b} C(w)$ 
for all $a,b \in \C$ and all $v,w \in V$. 
\item Let $V_1$ and $V_2$ be vector spaces with conjugations
$C_1$ and $C_2$.
A c-homomorphism, $h: V_1 \to V_2$, is a vector-space homomorphism that is compatible with the conjugations, i.e. $h \circ C_1 = C_2 \circ h$.
\item A test-function space is a complex separable locally-convex Hausdorff topological vector space on which a 
conjugation is defined. 
\item
A Q-map is a complex-linear map from a test-function space
into a $^\ast$-algebra, $\Phi: V \to P(\Phi)$, 
where $P(\Phi)$ is the polynomial algebra generated by the set of operators $\Phi(V)  
\cup \{1 \}$.\footnote{We note that if the field operators satisfy CCRs or CARs, then $P(\Phi)$ is generated by $\Phi(V)$ alone.}
$\Phi$ further satisfies $\Phi(f)^\ast = \Phi(Cf)$ for all $f \in V$. 
\item
A Q-theory is a pair $(\Phi, \omega)$, where $\Phi$ is a Q-map, $\omega$ is a state over $P(\Phi)$, and, for all $n \in \N$, 
\[
\omega \left( \prod_{m=1}^n \Phi( \, \cdot \, ) \right)
\]
is a multi-linear continuous functional on $V^n$.
\end{enumerate}
We note that we restrict test-function spaces to be separable and that larger test-function spaces will be considered in a subsequent paper. Moreover,
we will show in subsequent sections that quantum fields typically are Q-maps and that QFTs typically are Q-theories. We note however that Q-maps and Q-theories 
lack specific physical properties, like commutation relations, for example, and we emphasize that they serve a purely technical purpose in this paper.
In particular, the above definitions are not intended as a set of axioms for QFTs.

Let us further elaborate on our definitions. An example of a 
test-function space is Schwartz space, $S(\R^n)$. Complex
conjugation is given in $S(\R^n)$ by
\[
C( af + bg )(x) = \bar{a} \bar{f}(x) + \bar{b} \bar{g}(x)
= \bar{a} (Cf)(x) + \bar{b} (Cg)(x).
\]
Moreover, for any locally-convex topological vector space, $V$, a test-function space can be constructed as follows. Let $\bar{V}$ denote
the corresponding complex conjugate vector space (c.f. appendix A.2 in Ref. \cite{wal94}), and let $j: V \to \bar{V}$ denote the natural
anti-linear bijection. The set $\{ j(B): \, B \, \mathrm{open} \, \mathrm{in}\, V \}$ is a locally-convex Hausdorff topology on $\bar{V}$.
Let $W = V \oplus \bar{V}$
be equipped with the product topology of $V \times \bar{V}$, and define the conjugation $C(f,g) = (j^{-1}(g), j(f))$, then $W$ is a test-function space. 

A test-function space, $V$, always has an associated Q-map, which can be constructed as follows. Let
\[
\mathcal{A}_V = \bigoplus_{n=0}^\infty V^{\otimes n}
\]
denote the tensor algebra of $V$. For the sake of notational
convenience, we denote an element in $V^{\otimes n}$ by
\[
v_1 \otimes v_2 \otimes ... \otimes v_n = 
\bigotimes_{m=1}^n v_m 
\]
Moreover, we define the involution
\begin{eqnarray*}
C^{(n)} \left( \bigotimes_{m=1}^n v_m \right) & = & \, \bigotimes_{m=1}^m C(v_{n-m+1})
\quad ( v_m \in V ) \\
(a_n)^\ast & = & (C^{(n)}a_n) \quad ( (a_n) \in  \mathcal{A}_V),
\end{eqnarray*}
so that $\mathcal{A}_V$ is a (non-commutative)
$^\ast$-algebra. We further define the 
complex-linear Q-map, $\Phi_V: V \to \mathcal{A}_V$, by
\[
\Phi_V(v) = (0, v, 0, 0, ...) \quad (v \in V),
\]
then $P(\Phi_V) = \mathcal{A}_V$.
Note that for $V= S(\R^n)$, $\mathcal{A}_V$ is a BU-algebra. 

Multiple Q-maps can be combined into a single Q-map as follows.
Let $\Phi_i: V_i \to P(\Phi_i)$ ($i \in I$) be at most countably many Q-maps, for which multiplication and addition of operators are
defined. Define 
\[
V = \bigoplus_i V_i, \quad C (v_i) = (C_i v_i),
\]
and equip $V$ with the product topology, i.e. $V$ is a subspace of $\prod_i V_i$. 
$V$ is a complex separable locally-convex Hausdorff topological vector space, i.e. $V$ is a test-function space.
Define further the map
\[
\Phi((f_i)) = \sum_i \Phi_i(f_i), \quad
\Phi((f_i))^\ast = \sum_i \Phi_i(f_i)^\ast,
\]
then $\Phi$ is a Q-map.  \\[.15cm]
{\bf Proposition 1:} Let $\Phi: V \to P(\Phi)$ be a Q-map, let $W$ be a test-function space, and let $h: W \to V$
be a c-homomorphism, then $\Phi \circ h$ induces a $^\ast$homomorphism, $\pi: P(\Phi_W) \to P(\Phi)$,
so that $\Phi \circ h = \pi \circ \Phi_W$.\\[.15cm]
{\em Proof}:  $\Phi \circ h$ is a complex-linear function from
$W$ into $P(\Phi)$ that uniquely extends to an 
algebra homomorphism, $\pi$, 
from the tensor algebra $ \mathcal{A}_W$ of $W$ to $P(\Phi)$ by the
universal property of the tensor algebra. Let $C_V$ denote the conjugation on $V$, and let $C_W$ denote the conjugation on $W$.
$\pi$ further is a 
$^\ast$homomorphism:
\[
\pi((a_n) ) = a_0 + \sum_n 
\bigotimes_{m=1}^n \Phi( h( w_{m,n}))
\quad ( (a_n) \in \mathcal{A}_W, \, 
a_n = \bigotimes_{m=1}^n w_{m,n} \in W^{\otimes n}, 
\, w_{m,n} \in W ),
\]
and
\begin{eqnarray*}
\pi ( (a_n) )^\ast & = & \bar{a}_0 + \sum_n 
\bigotimes_{m=1}^n \Phi( h( w_{n-m+1,n} ))^\ast 
= \bar{a}_0 + \sum_n 
\bigotimes_{m=1}^n \Phi( C_V( h( w_{n-m+1,n} ) ) ) = \\
& = & \bar{a}_0 + \sum_n 
\bigotimes_{m=1}^n \Phi( h( C_W w_{n-m+1,n} ) ) 
= \pi ( (a_n)^\ast ).
\end{eqnarray*}
In particular, $( \Phi \circ h ) (w) ^\ast = ( \pi \circ \Phi_W) (w) ^\ast $ ($w \in W$). 
$\rule{5pt}{5pt}$ \\[.15cm]
{\bf Definition 1:} Let $\Phi_i: V_i \to P(\Phi_i)$ be Q-maps ($i = 1,2$). 
$\Phi_1$ is a core for $\Phi_2$ if there exists a continuous c-homomorphism, $h: V_1 \to V_2$, 
and a $^\ast$homomorphism, $\pi: P(\Phi_1) \to P(\Phi_2)$,
so that $h(V_1)$ is dense in $V_2$ and that  $\Phi_2 \circ h = \pi \circ \Phi_1$. If $h$ is surjective, then $\Phi_2$ is a quotient of $\Phi_1$.\\[.3cm]
{\bf Corollary 1:} Let $\Phi: V \to P(\Phi)$ be a Q-map, then $\Phi$ is a quotient of $\Phi_V$.\\[.15cm]
{\em Proof}:  The identity map, $\mathrm{id}: V \to V$, is a surjective continuous c-homomorphism, 
so that proposition 1 yields $\Phi = \pi \circ \Phi_V$. $\rule{5pt}{5pt}$ \\[.15cm]
As mentioned above, QFTs typically are Q-theories. Let us assume $\Phi_2 \circ h = \pi \circ \Phi_1$ as in definition 1, i.e. $\Phi_1$ is a core for $\Phi_2$, 
and let $(\Phi_2, \omega_2)$ be a QFT.  $\omega_1 = \omega_2 \circ \pi$ is a state on $P(\Phi_1)$, and  $(\Phi_1, \omega_1)$ is a Q-theory. 
We argue in the following that
$(\Phi_1, \omega_1)$ and $(\Phi_2, \omega_2)$ essentially yield the same quantum theory.
\subsection{Quantum theories from Q-map cores}
Let $(\Phi, \omega)$ be a Q-theory. $\omega$ induces the
representation of $P(\Phi)$ on a pre-Hilbert space, $D_\omega$,
by the GNS construction (for $^\ast$-algebras). Elements in $D_\omega$ are given by
equivalence classes of operators in $P(\Phi)$.
The representation is commonly denoted 
by $(H_\omega, \pi_\omega, \Omega_\omega)$, where $H_\omega$ is the completion of $D_\omega$, $\pi_\omega$ is a 
$^\ast$-homomorphism, and $\Omega_\omega$ is the unit vector 
corresponding to the unity operator, $1$. Expectation values
of operators in $P(\Phi)$ are given by
\[
\langle u_{[a]}, \pi(b) u_{[c]} \rangle =
\langle u_{[a]}, u_{[bc]} \rangle = \omega(a^\ast b c)
\quad (a,b,c \in P(\Phi)).
\]
The continuity property in our definition of a Q-theory further guarantees that 
\[
\langle u, \prod_{i_1}^n \Phi(f_i) v \rangle
\]
defines a complex-linear multi-continuous functional for all $u,v \in D_\omega$.
This is satisfied in scalar Wightman QFTs, for example.

However, let $\tau_\omega$ denote the
locally-convex topology generated by the set of semi-norms,
$\{ \| \pi_\omega(\, \cdot \, ) u \|: \, u \in D_\omega \}$,
on $P(\Phi)$.
Let us further assume that $V$ contains a  subset, $V_s$, so that $\pi_\omega(\Phi(f))$ is essentially self-adjoint
for all $f \in V_s$ and that $P(\Phi)$ is generated by $\Phi(V_s)$.
We call such a Q-theory regular. Let $\mathcal{A}$ be the C$^\ast$-algebra generated by the set $\{ \exp(i \pi_\omega(\Phi(f))): \, f \in V_s \}$,
then $\pi_\omega(P(\Phi))' = \mathcal{A}'$, i.e. the commutants agree, and $\pi_\omega(P(\Phi))'' = \mathcal{A}''$. 
$\mathcal{A}''$ may be seen as the algebra of
observables of the state $\omega$, which contains all 
projection-valued measures that are relevant in the respective theory (c.f. Def. 2.6.3 in Ref. \cite{bra79}). If the set $\{ \Phi(f): \, f \in V_s \}$
is irreducible, then the set $\{ \exp(i \pi_\omega(\Phi(f))): \, f \in V_s \}$ is also irreducible, and $\mathcal{A}''$ equals
the set of linear-bounded operators on $H_\omega$.
However, by von-Neumann's density theorem, $\mathcal{A}$ is dense in $\mathcal{A}''$ with respect to
the weak operator topology, and the restriction of any operator in $\mathcal{A}''$ to 
$D_\omega$ is the $\tau_\omega$-limit of a net of polynomials of operators in $\Phi(V)$. 

Let $\tilde{\Phi}$ be a core for $\Phi$, i.e. there exists a continuous c-homomorphism, $h$, and a $^\ast$-homomorphism, $\pi$, so that
$\Phi \circ h = \pi \circ \tilde{\Phi}$, then $\pi( P(\tilde{\Phi}))$ is dense in $P(\Phi)$
with respect to the $\tau_\omega$-topology. In particular,
each operator in $\pi_\omega ( P(\Phi))$ is the strong-graph
limit of a net of operators in $\pi_\omega \circ \pi( P(\tilde{\Phi}))$. If $(\Phi, \omega)$ is regular, then 
$\pi_\omega \circ \pi( P(\tilde{\Phi}))'' = \mathcal{A}''$.
\\[.3cm]
{\bf Proposition 2:} Let $(\Phi, \omega)$ be a Q-theory, and let $\tilde{\Phi}$ be a core for $\Phi$, then
there exists a $^\ast$-homomorphism, $\pi$, so that 
each operator in $\pi_\omega ( P(\Phi))$ is the strong-graph
limit of a net of operators in $\pi_\omega \circ \pi( P(\tilde{\Phi}))$.\\[.15cm]
Let us re-formulate proposition 2 into a looser statement: 
$(\Phi, \omega)$ and $(\tilde{\Phi}, \omega \circ \pi)$ essentially yield the same 
theory, which emerges from a representation of $P(\tilde{\Phi})$. 
\subsection{A universal Q-map}
Let $V_0$ be the vector space of cofinite complex sequences,
\[
V_0 = \bigoplus_{n = 1}^\infty \, \C,
\]
and define the conjugation
\[
C_0( (c_n) ) = ( \bar{c}_n ) \quad ( (c_n) \in V_0).
\]
{\bf Lemma 1:} Let $V$ be a test-function space. There exists a c-homomorphism, $h: V_0 \to V$, so that
$h(V_0)$ is dense in $V$. If $V$ is finite-dimensional, then $h$ is surjective. \\[.15cm]
{\em Proof}: Let $\{ f_n \}$ be a countable, dense subset of $V$, let $W$ be the linear span of $\{ f_n \}$, and
let $C$ denote the conjugation on $V$.
Let further $\{ e_n \}$ be a maximal linear-independent subset of 
\[
\{ (f_n + C(f_n) \} \cup \{ i( f_n - C(f_n) ) \} ,
\]
then $\{ e_n \}$ is a basis of $W$. Note that if $W$ is finite-dimensional, then $W$ is closed and thus $W = V$.
Moreover, the function
\[
h( (c_n) ) = \sum_{n=1}^{\dim W} c_n e_n \quad ( (c_n) \in V_0)
\]
is a vector-space homomorphism between $V_0$ and $W$, which is compatible with the conjugations,
\[
C( h( (c_n) ) ) = \sum_n \bar{c}_n e_n = h( C_0( (c_n) ) ). \quad \rule{5pt}{5pt}
\]
Let $\mathcal{V}$ be the class of test-function spaces, and for $V  \in \mathcal{V}$ let $h_V: V_0 \to V$ be a  
c-homomorphism so that $h(V_0)$ is dense in $V$. If $V = V_0$ as sets, and if the conjugation on $V$ is $C_0$,
then we choose $h_V = \mbox{id}$. We equip $V_0$ with the initial topology generated by the set
$\{ h_V^{-1}(B): \, B \, \mathrm{open} \, \mathrm{in}\, V, \, V \in \mathcal{V} \}$. Note that this is the weakest topology
for which all functions $h_V$ ($V \in \mathcal{V}$) are continuous.\\[.3cm]
{\bf Lemma 2:} $V_0$ is a test-function space.  \\[.15cm]
{\em Proof}: We first note that a function $f: Z \to V_0$ is continuous if and only if $h_V \circ f$ is continuous for all $V \in  \mathcal{V}$. Let $f_{c, W}$ denote
multiplication with $c \in \C$ on a vector space, $W$. $f_{c, V_0}$ is continuous since
$h_V \circ f_{c, V_0} = f_{c, V} \circ h_V$, and since scalar multiplication is continuous on $V$ for all $V \in  \mathcal{V}$.
Let $g_{W}(a,b) = a + b$ denote the addition function on a vector space, $W$. $g_{V_0}$ is continuous since
$(h_V \circ g_{V_0}) (a,b) = g_V( h_V(a), h_V(b) ) = h_V(a) + h_V(b)$, and since addition is continuous on $V$ for all $V \in  \mathcal{V}$.
Let further $a,b \in V_0$, $a \neq b$, let $V = V_0$ as sets, let $V$ be equipped with the topology induced by the semi-norms $p_m( (c_n) ) = | c_m|$ 
($ (c_n) \in V$), and let $C_0$ be the conjugation on $V$, then $V$ is a test-function space. The topology of $V$ is contained in the topology of $V_0$
by definition of $V_0$.  
Since $V$ is Hausdorff, there exist open sets $A$ and $B$ in $V$ so that 
$a \in A$, $b \in B$, and $A \cap B = \emptyset$. $V_0$ is therefore a Hausdorff space since $A$ and $B$ are also open in $V_0$.
For each $V \in  \mathcal{V}$ let $\mathcal{B}_V$ denote a neighborhood base of $0$ of balanced, convex, absorbing sets, and let $\mathcal{B}_{V_0}$
be the set of finite intersections of sets in $\{ h_V^{-1}(B): \, B \in \mathcal{B}_V, \, V \in \mathcal{V} \}$. $\mathcal{B}_{V_0}$ is a neighborhood base
of $0$. Due to linearity, $h_V^{-1}(B)$ is a balanced, convex, and absorbing set for all $B \in \mathcal{B}_V$ and all $V \in \mathcal{V}$. Let
$B  \in \mathcal{B}_{V_0}$, $B = C_1 \cap ... \cap C_n$, and let $a \in V_0$. $B$ is balanced and convex. For each $C_i$ ($1 \leq i \leq n$) there exists a $t_i > 0$
so that $a \in t C_i$ if $t \geq t_i$. Let $t_0 = \max \{ t_1, ..., t_n \}$, then $a \in tB$ if $t \geq t_0$. $B$ is therefore absorbing, and $V_0$ is locally convex.
Moreover, $V_0$ is the union of countably many finite-dimensional spaces,
\[
V_0 = \bigcup_n V_{0,n}, \quad V_{0,n} = \{ (c_m) \in V_0: \, c_m = 0 \, \forall \, m > n \}.
\]
Since each $V_{0,n}$ is finite-dimensional, the respective subspace topologies are equivalent to the Eucledian topologies, which entails that each
$V_{0,n}$ is separable. Let $W_{0,n}$ be a countable, dense subset of $V_{0,n}$, then $\bigcup_n W_{0,n}$ is a countable, dense subset of $V_0$.
\quad \rule{5pt}{5pt}\\[.15cm]
Let $\Phi_0: V_0 \to \mathcal{A}_0$ be the Q-map associated with $V_0$.
Lemma 1 and proposition 1 yield the following theorem.\\[.3cm]
{\bf Theorem 1:} Let $\Phi: V \to P(\Phi)$ be a Q-map, then $\Phi_0$ is a core for $\Phi$.
If $V$ is finite-dimensional, then $\Phi$ is a quotient of $\Phi_0$.\\[.15cm]
Considering proposition 2, $\Phi_0$ actually is a universal Q-map, since any Q-theory basically is a Q-theory of $\Phi_0$, which emerges
from a representation of $P(\Phi_0)$.\\[.3cm]
{\bf Theorem 2:} Let $\omega$ be a state over $P(\Phi_0)$, then $(\Phi_0,\omega)$ is a Q-theory.\\[.15cm]
{\em Proof}: We need to show that, for all $n \in \N$, 
\[
\omega^{(n)} = \omega \left( \prod_{m=1}^n \Phi_0( \, \cdot \, ) \right)
\]
is a multi-linear continuous functional on $V_0^n$. Let $F$ be the set of linear functions from $V_0$ to $\C$. For each $f \in F$ we define
the semi-norm $p_f(v) = |f(v)|$ ($v \in V_0$). The set of semi-norms, $\{p_f\}_{f \in F}$, defines a locally-convex Hausdorff topology on the set $V_0$.
Let $V_0^F$ denote the corresponding topological space. We choose $C_0$ as conjugation on $V_0^F$, so that $V_0^F$ is a test-function space.
Each $f \in F$ is continuous when considered as a function from $V_0^F$ to $\C$. Due to the definition of the test-function space $V_0$, each open set
in $V_0^F$ is also open in $V_0$, so that each $f \in F$ is also continuous when considered as a function from the test-function space $V_0$ to $\C$.
Hence, $\omega^{(n)} $ is continuous in each argument, and therefore it is continuous on $V_0^n$.
\quad \rule{5pt}{5pt}\\[.15cm]
Let $S_0$ be the set of states over $P(\Phi_0)$. Each $a \in P(\Phi_0)$ defines a linear functional on $S_0$ by $l_a(\omega) = \omega(a)$
for $\omega \in S_0$. The corresponding set of semi-norms, $p_a(\omega) = | l_a(\omega) | = | \omega(a) |$, defines a topology on $S_0$,
and a net $(\omega_i)$ in $S_0$ converges to an $\omega \in S_0$ with respect to that topology, 
if $\lim_i \omega_i(a) = \omega(a)$ for all $a \in P( \Phi_0)$.

Let  $V_{0,n} = \{ (c_m) \in V_0: \, c_m = 0 \, \forall \, m > n \}$. We will argue in Sec. \ref{LQFT} that $V_{0,n}$ is the test-function space of a
quantum system with $n$ degrees of freedom. Such quantum systems typically are considered in lattice QFTs. However, 
let $\Phi_{0,n}$ denote the corresponding Q-map, and let
$P_n: V_0 \to V_{0,n}$ be the canonical projection, i.e. $P_n( (c_m) ) = (c_1, ..., c_n, 0, 0, ...)$ for $(c_m) \in V_0$, then 
$\Phi_{0,n} = \Phi_0 \circ P_n$ and $P(\Phi_{0,n})$ is a sub-algebra of $P(\Phi_0)$. Since $P_n$ is a c-homomorphism, there exists a 
corresponding $^\ast$homomorphism, $\pi_n: P(\Phi_0) \to P(\Phi_{0,n})$, by proposition 1 so that $\Phi_{0,n} = \pi_n \circ \Phi_0$. 
Let $(\Phi_0,\omega)$ be
a Q-theory, then $\omega_{(n)} = \omega |_{P(\Phi_{0,n})}$ defines a state on $P(\Phi_{0,n})$, and $(\Phi_0,\omega_{(n)} \circ \pi_n )$ is a Q-theory.
For the sake of notational convenience we denote $\omega_{(n)} \circ \pi_n$ simply by $\omega_{(n)}$ in the following.
$(\Phi_0,\omega_{(n)})$ represents a reduced system. We note that the 
series $( \omega_{(n)} )$ converges to $\omega$, and that the QFTs discussed in this paper can therefore be approximated
on lattices, see also Sec. \ref{LQFT}. \\[.3cm]
{\bf Theorem 3:} Let $(\omega_i)$ be a net of states over $P(\Phi_0)$. $(\omega_i)$ converges to a state $\omega$ if and only if
$(\omega_{i,(n)})$ converges to $\omega_{(n)}$ for all $n \in \N$.\\[.15cm]
{\em Proof}: $\omega_{i,(n)} \to \omega_{(n)}$ is a consequence of $\omega_i \to \omega$. Let us assume that $\omega_{i,(n)} \to \omega_{(n)}$
for all $n \in \N$, and let $a \in P(\Phi_0)$. Since $a \in P(\Phi_{0,n})$ for some $n \in \N$, and since for $n < m$, $V_{0,n} \subset V_{0,m}$ and 
$\omega_{(n)} = \omega_{(m)} |_{P(\Phi_{0,n})}$,
the states $\omega_{(n)}$ defines a unique state $\omega$ on $P(\Phi_0)$. \rule{5pt}{5pt}\\[.15cm]
In the remaining part of this paper, we will show that theorem 1 applies to many well-known QFTs, and that theorems 2 and 3 define the continuum limit of lattice
QFTs in our approach.
\section{Application to Wightman quantum field theories}

The relation of our approach to Wightman QFTs can be conveniently
discussed with the help of the Wightman reconstruction theorem. 
For the sake of convenience,
let us consider a hermitian scalar Wightman QFT. 
In the reconstruction theorem, Wightman QFTs are recovered as
representations of a Borchers-Uhlmann algebra \cite{str00}.
The test-function space in Wightman QFTs is Schwartz space, $S(\R^d)$ ($d \geq 2$). Using the terminology and the definitions of
Sec. \ref{GENERAL}, the Borchers-Uhlmann algebra is given by $\mathcal{A}_V$ with $V = S(\R^d)$, and the
corresponding Q-map is denoted by
$\Phi_V: V \to \mathcal{A}_V$. Note that $S(\R^d)$ is separable, and that corollary 1 and theorem 1 apply.\\[.3cm]
{\bf Corollary 2:} Let $(\Phi, \omega)$ be a hermitian scalar Wightman QFT, then $\Phi$ is a quotient of  $\Phi_V$ ($V = S(\R^d)$), and 
$\Phi_0$ is a core for $\Phi$.\\[.15cm]
We note that corollary 2 applies to any (hermitian scalar)
Wightman QFT involving $d \geq 2$ space-time dimensions, and to 
Wightman QFTs of interacting quantum fields as well as free quantum fields.
Moreover, considering proposition 2, we can re-formulate corollary 2 into a looser statement: 
There exists a $^\ast$-homomorphism, $\pi$, so that
$(\Phi, \omega)$ and $(\Phi_0, \omega \circ \pi)$ essentially yield the same 
quantum theory, which emerges from a representation of $P(\Phi_0)$.

Let us discuss two examples of Wightman QFTs in more detail.
In general, the set of field operators in Wightman QFTs is 
irreducible. Moreover, let $V_s$ denote the subset of 
real functions in $S(\R^d)$, then the set
$\{ \Phi(f): \, f \in V_s \}$ is also 
irreducible, and $\Phi(V_s)$ generates $P(\Phi)$.
In the usual Fock-space representation of 
free scalar fields \cite{ree75}, the operators $\Phi(f)$ ($f \in V_s$) are
essentially self-adjoint, i.e. the QFT is regular. 
Unfortunately the situation is less
straightforward for Wightman QFTs of interacting fields, since
there do not exist that many examples. However, let us consider
$P(\varphi)_2$ as presented in Ref. \cite{gli71}.

$P(\varphi)_2$ is defined in flat space-time with one time 
dimension and one space dimension. 
The corresponding Fock space, $\mathcal{F}$, 
of the free hermitian scalar QFT is the symmetric tensor
algebra over $L_2(\R)$. For each open bounded interval,
$B \subset \R$, let $\mathcal{A}(B)$ denote the von-Neumann
algebra generated by the operators $\exp(i \varphi(f_1) + 
i \pi(f_2))$ ($f_1, f_2 \in C^\infty_0(B)$, $f_1, f_2$ real) and let 
$\mathcal{A}$ denote the norm closure of 
$\bigcup_B \mathcal{A}(B)$. $P(\varphi)_2$ is
constructed by considering the GNS representation,
$(H_\omega, \pi_\omega, \Omega_\omega)$, of $\mathcal{A}$
with respect to a specific state, $\omega$. In this representation,
the unitary groups 
\[
W_t(f_1, f_2) = \pi_\omega( 
\exp(i t \varphi(f_1) + i t \pi(f_2)) ) \quad (
f_1, f_2 \in C^\infty_0(B))
\]
are strongly continuous, and they have self-adjoint generators.

Let $\Phi(f) = \varphi(\mathrm{Re}(f)) + i \pi(\mathrm{Im}(f))$, then $\Phi$ is a Q-map.
Let $\Phi'$ be the restriction of $\Phi$ to $ C^\infty_0(\R)$
(assuming the Schwartz-space topology), then $\Phi'$ is a core
for $\Phi$, and $(\Phi',\omega)$ is a regular Q-theory. 
Also, let $W_s = V_s \cap C^\infty_0(\R)$, then the set
$\{ \Phi(f): \, f \in W_s \}$ is irreducible, i.e. $ \mathcal{A} \subset P(\Phi')'' = B(H_\omega)$.
We can therefore say that
$(\Phi, \omega)$  and $(\Phi', \omega)$ essentially yield the same quantum theories.
Moreover, $\Phi_0$ is a core for $\Phi'$, and proposition 2 applies.

\section{Application to free quantum field theories}
\label{freeQFT}

In this section, we discuss the relation of our general approach to free
QFTs that implement canonical commutation relations (CCRs) or canonical anti-commutation relations (CARs) on Fock space.
We first discuss both cases together without specifying if the Fock space, $\mathcal{F}$, is symmetric or anti-symmetric. We assume
however that $\mathcal{F}$ is constructed over an infinite-dimensional, separable Hilbert space, $\mathcal{H}$. 

In conventional representations \cite{bra81}, the annihilation and creation operators, $a(f)$ and $a^\ast(f)$, are both defined over $\mathcal{H}$, so that
$a(f)$ is complex anti-linear and that $a^\ast(f)$ is complex linear. We note that $a(f)$ and $a^\ast(f)$ are densely defined, closed, and that $a(f)^\ast = a^\ast(f)$. 
If there is a complex conjugation, $C$, defined on $\mathcal{H}$, then 
$\mathcal{H}$ and $\mathcal{H}^2$ are test-function spaces, and we can introduce the complex-linear Q-map
$\Phi_1 (f,g)  = a^\ast(f) + a(Cg)$. Since 
$\Phi_1(f,0) = a^\ast(f)$ and
$\Phi_1(0,Cf) = a(f)$, 
$P(\Phi_1)$ is the polynomial algebra generated by the irreducible set of operators $A = 
\{   a(f), a^\ast(f) \}_{f \in \mathcal{H}} $. In particular, in the symmetric case (CCRs), we obtain
\[
\Phi(f) = \frac{ a^\ast(f) + a(f) }{\sqrt{2}} = \frac{ \Phi_1(f,0) + \Phi_1(0,Cf) }{\sqrt{2}}.
\]
Since $\mathcal{H}^2$ is separable, we can apply theorem 1.\\[.3cm]
{\bf Corollary 3:} $\Phi_0$ is a core for $\Phi_1$.\\[.15cm]
The annihilation operators in free QFTs can conveniently be chosen as complex-linear operator-valued functionals if they are defined
over the dual Hilbert space instead, i.e. one considers $a(f)$ with $f \in \mathcal{H}^\ast$ \cite{wal94}. One advantage of this choice is that vacuum expectation values
become multi-linear functionals over the test-function space. However, by the Riesz lemma, there is a natural complex anti-linear bijection, 
$j: \mathcal{H} \to \mathcal{H}^\ast $, and we can define the corresponding complex conjugation on $W = \mathcal{H} \otimes \mathcal{H}^\ast$ by
$C(f,g) = (  j^{-1} (g), j(f) )$, i.e. $W$ is a test-function space (c.f. Sec. \ref{GENERAL}). 
In particular, let $W_2 = l_2(\N) \otimes l_2(\N)^\ast$, and let $\Phi_2 = \Phi_{W_2}$. 
For a separable Hilbert space, $\mathcal{H}$, choose a unitary
operator $U : l_2(\N) \to \mathcal{H}$, let $\bar{U}: \mathcal{H}^\ast \to l_2(\N)^\ast$ denote the corresponding dual unitary operator,
define $\Phi_2'(f,g)  = a^\ast(f) + a(g)$, and define $\pi_U( \Phi_2(f,g)) = \Phi_2'(Uf,\bar{U} g)$. $\pi_U$ is a $^\ast$-homomorphism
between $P(\Phi_2)$ and $P(\Phi_2')$, $\Phi_2'$ is a quotient of $\Phi_2$, and 
the Fock representation is a $^\ast$-homomorphic representation of $P(\Phi_2)$. We summarize.\\[.3cm]
{\bf Corollary 4:} In any free QFT, the polynomial algebra generated by the irreducible set of operators 
$\{   a(f), a^\ast(g) \}_{f \in \mathcal{H}, g \in \mathcal{H}^\ast} $ is $^\ast$-homomorphic
to $P(\Phi_2)$, and  $\Phi_0$ is a core for $\Phi_2$.\\[.15cm]
Adopting the same loose language as after proposition 2, we state that free QFTs emerge from Fock representations of $P(\Phi_0)$.

Let us discuss two examples of free QFTs. In Ref. \cite{dim80},
CCRs are defined with $\mathcal{H} = L_2(S)$, where $S$ is a Cauchy surface of a globally hyperbolic manifold. 
Independence of the actual choice of $S$ is
due to a $^\ast$-isomorphism between representations of the CCRs over the same vector space ($C_0^\infty(M)$). We note that a 
more detailed account is given in Ref. \cite{wal94}, where especially the arbitrariness of the choice of scalar product 
in the definition of $\mathcal{H}$ is discussed. However, we note that corollary 3 applies irrespective of the specifically chosen
globally hyperbolic manifold and background metric, i.e. 
the QFTs emerge from representations of $P(\Phi_0)$ and $P(\Phi_2)$, respectively.

The second example are Dirac quantum fields on a globally 
hyperbolic manifold, $M$, that are constructed in Ref.  \cite{dim82}. The construction 
is based on the definition of a scalar product on
$C_0^\infty(DS)$, where $DS$ is the
Dirac spinor bundle of spinors on a Cauchy surface, $S$. $DS$
is a vector bundle, and $C_0^\infty(DS)$ is locally isomorphic
to $C_0^\infty(S)^4$. The completion of $C_0^\infty(DS)$ yields
a separable Hilbert space, $\mathcal{H}$. The dual Hilbert space, $\mathcal{H}^\ast$, is the closure of 
$C_0^\infty(D^\ast S)$, the space of cross sections with compact support over the dual vector bundle $D^\ast S$.
The representation of CARs over $S$ is further defined as a representation of the CARs over the pair $\mathcal{H}$, $\mathcal{H}^\ast$.
We note that corollary 4 applies irrespective of the specifically chosen
globally hyperbolic manifold and background metric, i.e. 
the QFTs emerge from representations of $P(\Phi_0)$ and $P(\Phi_2)$, respectively.

\section{Application to perturbative quantum field theory in curved space-times} 

The perturbative formulation of quantum theories of interacting 
fields in curved space-time is closely related to recent achievements in the field
of algebraic QFT \cite{hol10,bru03}.
In conventional approaches (c.f. Sec. \ref{freeQFT}), free
quantum fields in curved space-time are operator-valued 
distributions over a space of smooth, compactly-supported test 
functions on a specific space-time. Independence of the specific space-time background can be further
achieved in a categorial approach \cite{bru03}, which is also applicable to
perturbatively interacting quantum fields. Starting from free QFTs,
renormalized perturbation theory on curved space-times was explicitely formulated by methods of microlocal analysis \cite{bru00,hol01,hol02}.

Let us consider the example of a QFT of scalar hermitian fields as it is presented in Ref. \cite{hol01}.
The first step is to enlarge the quantum-field algebra of the corresponding free QFT with the help of microlocal analysis. 
The test function space of a free scalar hermitian QFT on a 
globally hyperbolic space-time manifold, $M$, is $C_0^\infty(M)$.
Assuming a quasi-free Hadamard state, we can represent the 
the quantum-field algebra by the GNS construction. In such a
representation, however, the quantum fields are operator-valued
distributions over a larger, distributional test-function space. 
Let 
\[
W_n(x_1, .., x_n) = \, :\phi(x_1) ... \phi(x_n):_\omega
\]
denote the (Wick-ordered) operator-valued distribution that is defined 
over $C_0^\infty(M^n)$ and let $\omega$ denote a quasi-free
Hadamard state ($n \geq 1, \, W_0 = 1$). As can be shown by
microlocal analysis, the operator-valued distributions are 
defined on a larger space, $E_n'$, which contains
$C_0^\infty(M^n)$ and which is a subspace of 
the dual space $C_0^\infty(M^n)'$. Distributions in $E_n'$ are
compactly supported, and they satisfy the wave-front condition
$\WF(t) \subset G_n$ ($t \in E_n'$), 
where $G_n = (T^\ast M)^n \setminus H_n$ and
\[
H_n = \{ (x,k) : \, x \in M, \, k \in (\bar{V}^+)^n \cup 
(\bar{V}^-)^n \}
\]
Note that such distributions can be multiplied 
with each other, so that local Wick polynomials can 
rigorously be defined. Let $\mathcal{W}$ be the $^\ast$-algebra
of operators generated by 1 and elements $\{W_n(t) \}_{n \in \N, t \in E_{n}}$ (c.f definition 2.1 in Ref.
\cite{hol01}). Note also that 
$E_n'$ does not have a proper topology in the sense of 
H\"ormander \cite{hor85}, but that it rather has a 
so-called pseudo topology, see Ref. \cite{hol01} for
more details.

However, let us endow $E_n'$ with the sub-space topology, 
which is inherited from
the weak$^\ast$ topology on $C_0^\infty(M^n)'$.
$E_n'$ is separable, and complex conjugation is well-defined since $C_0^\infty(M^n)$ is a dense subspace, 
i.e. $E_n'$ is a test-function space 
according to our definition in Sec. \ref{GENERAL}. Each
$W_n$ is a Q-map, and the corresponding polynomial algebra, $P(W_n)$,
is a sub-algebra of $\mathcal{W}$.
As outlined in Sec. \ref{GENERAL}, 
we can combine the Q-maps by defining the test-function space
$E' = \bigoplus_n E_n'$ and the Q-map
$\Phi_3 = \sum_n W_n$, so that 
we obtain $P(\Phi_3) = \mathcal{W}$. 
We apply theorem 1.\\[.3cm]
{\bf Corollary 5:} $\Phi_0$ is a core for $\Phi_3$.\\[.15cm]
In the perturbative approach, interacting fields are 
formally defined by a perturbation series. Series
of operators in $\mathcal{W}$ form an algebra
$\mathcal{X} = \mathcal{W}^{\N}$ with multiplication
$(a_n) \star (b_n) = (a_0 b_0, a_1 b_0 + a_0 b_1, ...)$.
Note that the product is defined as if one formally
multiplies $\sum_n a_n$ and $\sum_n b_n$. Let us
consider the test-function space $\bigoplus_n E'$, on which
we define the Q-map $\Phi_4(f) = (\Phi_3(f_n))$. Using
the CCRs, one can further show that $P(\Phi_4)$ contains all co-finite sequences in $\mathcal{X}$. Note that the co-finite sequences are
dense in $\mathcal{X}$ if we consider a weak topology on $\mathcal{W}$ and the product topology on $\mathcal{X}$. However, we can again
apply theorem 1.\\[.3cm]
{\bf Corollary 6:} $\Phi_0$ is a core for $\Phi_4$.\\[.15cm]
Moreover, the interacting-field algebra  is defined in Ref. \cite{hol03}
as a sub-algbera of $\mathcal{X}$ as follows. There exists a multi-linear map
\[
T_{L_1}^{(n)}:  \mathcal{D}_1(M,\mathcal{V})^n \to \mathcal{X},
\]
where $\mathcal{D}_1(M,\mathcal{V})$ is a vector space and $L_1$ denotes the Lagrangian.
$\mathcal{V}$ is a vector space, which is generated by a countably infinite (Hamel) basis,
and $\mathcal{D}_1(M,\mathcal{V})$ is the space of compactly-supported smooth densities on $M$ with values in $\mathcal{V}$. An element
$F \in \mathcal{D}_1(M,\mathcal{V})$ can be uniquely expressed as a finite sum, $F = \sum f_i v_i$, where $f_i \in C_0^\infty(M)$ and
$v_i \in \mathcal{V}$. As a vector space, $\mathcal{D}_1(M,\mathcal{V})$  is therefore isomorphic to $\bigoplus_n C_0^\infty(M)$. The interacting-field algebra
is further defined as the algebra generated by the images of the maps $T_{L_1}^{(n)}$. 

Let us endow $\bigoplus_n C_0^\infty(M)$ with the topology induced by  $C_0^\infty(M)^{\N}$, and let the conjugation on $\bigoplus_n C_0^\infty(M)$
be induced by the usual complex conjugation on $C_0^\infty(M)$. $ \mathcal{D}_1(M,\mathcal{V})^n$ and $(\bigoplus_m C_0^\infty(M))^n$ are hence 
test-function spaces for all $n \in \N$, and
the maps $T_{L_1}^{(n)}$ satisfy our definition of a Q-map, see Sec. \ref{GENERAL}. Let $\Phi_5$ denote the Q-map combining the
$T_{L_1}^{(n)}$, then the interacting-field algebra is $P(\Phi_5)$. We can again apply theorem 1.\\[.3cm]
{\bf Corollary 7:} $\Phi_0$ is a core for $\Phi_5$.\\[.15cm]
Moreover, in Ref. \cite{hol07}, an algorithm is
presented to construct the Wilson operator-product expansion (OPE). The algorithm is generally applicable to perturbative interacting QFT in 
Lorentzian curved space-times, and it is explicitly presented for the example of a scalar hermitian self-interacting field. However, 
as proposed in Ref. \cite{hol10}, the OPE can actually be elevated to a fundamental level, so that the
QFT is determined by its OPE. This yields a general axiomatic framework for QFTs in curved space-times. In particular, the algebra of 
interacting quantum fields, $\mathcal{F}_i$, is obtained by factoring the corresponding free algebra, $\mathcal{F}_0$, by a set of relations arising from properties of the OPE
coefficients. These relations define an ideal, $I$, in $\mathcal{F}_0$, and $\mathcal{F}_i = \mathcal{F}_0 / I$. 
We note that the corresponding quotient map, $\pi_I: \mathcal{F}_0 \to \mathcal{F}_i$ is a $^\ast$homomorphism. As pointed out in this paper so far, 
we typically can find a $^\ast$homomorphism, $\pi_0: P(\Phi_0) \to \mathcal{F}_0$, so that $\pi_0(P(\Phi_0))$ is dense in $\mathcal{F}_0$ with respect to an appropriate
topology. Then, $\pi_I \circ \pi_0$ is a  $^\ast$homomorphism that maps $P(\Phi_0)$ onto a dense set in $\mathcal{F}_i$ with respect to another topology. 
Adopting the same loose language as after proposition 2, we then can state that a perturbative interacting  QFT emerges from a representation of $P(\Phi_0)$,
if the corresponding free QFT does.

\section{Further representations of $P(\Phi_0)$}

\subsection{Lattice quantum field theory}
\label{LQFT}

Let us use quantum chromodynamics on a lattice (LQCD) as an example of a lattice QFT.
LQCD is a non-perturbative approach to QCD. Calculations usually are performed 
using the Feynman path-integral approach. Starting point is a set of CCRs and CARs \cite{wei95} for
symmetric operators $x_a$, $p_a$, $\tilde{x}_c$, and $\tilde{p_c}$:
\begin{eqnarray*}
[ q_a, p_b ] & = & i \delta_{a,b} \, , \\
\! [ q_a, q_b ] & = &  [ p_a, p_b ] = 0 \, , \\
\{ \tilde{q}_c, \tilde{p}_d \} & = & i \delta_{c,d} \, , \\
\{ \tilde{q}_c, \tilde{q}_d \} & = & \{ \tilde{p}_c, \tilde{p}_d \} = 0 \, .
\end{eqnarray*}
The indexes represent the degrees of freedom of the quantum system, and they consist of a position, $x$, and a field index. 
In conventional QFT, $x$ formally is continuous, but in LQCD, $x$ is discrete and only takes finitely many values.
In particular, the Feynman integrals are rigorously defined in LQCD, and calculations are performed in exactly
the same way as in conventional QFT, where a space-time continuum is considered \cite{gup98}.

Let us assume that $a$ and $c$ take finitely many value, i.e. $1 \leq a \leq m$ and $1 \leq c \leq n$.
Let $\mathcal{A}_{m,n}$ be the operator algebra generated by $\{ x_a, p_a, \tilde{x}_c, \tilde{p_c} : \, 1 \leq a \leq m, \, 1 \leq c \leq n \}$. 
$\mathcal{A}_{m,n}$ is the operator algebra of the specific LQCD model instance. 
We define a function $x: \C^m \to \mathcal{A}_{m,n}$ by $x( (\delta_{ab})_{1 \leq b \leq m}) = x_a$ and by complex-linear
extension, and we define $x( (z_a) )^\ast = x( (\bar{z}_a) )$. $x$ is a Q-map, and we analogously define the Q-maps
$p$, $\tilde{x}$, and $\tilde{p}$. We further combine these Q-maps into one Q-map, $\Phi_{6,m,n}: V_{m,n} \to \mathcal{A}_{m,n}$, 
$V_{m,n} = \C^m \otimes \C^m \otimes \C^n \otimes \C^n = \C^{2m + 2n}$. Since $V_{m,n}$ is finite-dimensional, we can apply theorem 1.\\[.3cm]
{\bf Corollary 8:} $\Phi_{6,m,n}$ is a quotient of $\Phi_0$.\\[.15cm]
In the continuum limit, the lattice spacing is supposed to approach 0, i.e. the indexes $a$ and $c$ take an increasing number of values ($n,m \to \infty$). 
We note that this does not lead to truly continuous
indexes, and that continuous indexes must be differently treated. 
However, let us assume that for each grid size a state $\omega_{m,n}$ is determined. Since $\Phi_{6,m,n}$ is a quotient of $\Phi_0$,
we obtain a series of states over $P(\Phi_0)$. Let us denote the series by $(\omega_j)$ for the sake of notational convenience. Since 
$V_{m,n} = V_{0,2m+2n}$ and since $\omega_{j,(k)} = \omega_{j,(2n+2m)} |_{P(\Phi_{0,k})}$ ($k \leq 2n+2m$), we obtain a series of
states, $\omega_{j,(k)}$, on each sub-algebra $P(\Phi_{0,k})$ of $P(\Phi_0)$. If each series, $(\omega_{j,(k)})$, converges on the
corresponding sub-algebra $P(\Phi_{0,k})$, then there exists a unique limit state $\omega$ by theorem 3, and $(\Phi_0, \omega)$ is a Q-theory
by theorem 2. We note however that $(\Phi_0, \omega)$ is not necessarily a QFT, since one has to additonally ensure that essential physical features are
retrieved in that limit.

Let us rephrase this result: Increasing grid sizes yield a series of states. Each state of a specific grid defines a so-called reduced state on each sub-grid. 
A necessary and sufficient criterion for a unique limit state in the continuum limit is that the series of reduced states converges on each sub-grid. 
If a limit state exists, then we obtain a Q-theory as defined in this paper, and if appropriate physical features are retrieved then the Q-theory is also a 
reasonable QFT.

\subsection{String theory}

We base our discussion of string theory on the lecture notes of R. J. Szabo \cite{sza02}. 
String theory is still work in progress, and a thorough discussion of the relation of string theory
to the approach in this paper is elusive so far. However, there are five different consistent formulations of string theory that are commonly seen as perturbative 
expansions of a unique underlying theory (M-theory), which is however not well understood yet. The five theories are related by dualities that map perturbative
states in one theory to non-perturbative states in another theory.

However, the quantization of the bosonic string yields a countable set of raising and lowering operators that satisfy the relation $(a^\mu_n)^\ast = a^\mu_{-n}$,
and, if closed strings are considered,
 $(\tilde{a}^\mu_n)^\ast = \tilde{a}^\mu_{-n}$ ($0 \leq \mu \leq d, \, n \in \N$). These operators satisfy CCRs. There are also zero-mode operators,
$x^\mu_0$ and $p^\mu_0$, that are conjugate to each other and that also satisfy the CCRs. The operators act on a Fock space, and we can combine them
into an equivalent set of self-adjoint operators as follows:
\begin{eqnarray*}
x^\mu_n & = & \frac{a^\mu_n + a^\mu_{-n}}{2}, \quad p^\mu_n = \frac{i(a^\mu_n - a^\mu_{-n})}{2} \\
\tilde{x}^\mu_n & = & \frac{\tilde{a}^\mu_n + \tilde{a}^\mu_{-n}}{2}, \quad \tilde{p}^\mu_n = \frac{i(\tilde{a}^\mu_n - \tilde{a}^\mu_{-n})}{2} .
\end{eqnarray*}
We assume that observables in bosonic string theory are contained in the closure of the operator algebra, $\mathcal{A}_b$, 
generated by the operators $\{ x^\mu_n, p^\mu_n, \tilde{x}^\mu_n, \tilde{p}^\mu_n \}_{n \in \N}$ with respect to an appropriate topology. However,
let $v_n = (\delta_{n,m})_{m \in \N}  \in V_0$ and let $x^\mu(v_{n+1}) =  x^\mu_n$ and $p^\mu(v_{n+1}) =  p^\mu_n$ for $n \in \N_0$, then, by complex-linear
continuation,  $x$ and $p$ define Q-maps over the test-function space $V_0$. Analogously we can define the additional Q-maps 
$\tilde{x}$ and $\tilde{p}$ if we consider closed strings. 

Fermions further need to be included in string theory to avoid
inconsistencies. 
Canonical quantization yields another countable set of
operators satisfying $(\psi^\mu_r)^\ast = \psi^\mu_{-r}$ ($r = 0, \pm1, \pm \frac{1}{2}, \pm 2, \pm \frac{3}{2}, ...$). These operators generate the
corresponding operator algebra  $\mathcal{A}_f$, and we simply denote the total operator algebra generated by operators in 
$\mathcal{A}_b$ and $\mathcal{A}_f$ by $\mathcal{A}$. The Hilbert spaces of the various string theories are subspaces of 
the full Hilbert space that is obtained by canonical quantization,
and there are corresponding representations of $\mathcal{A}$
on these sectors. As for the bosonic case, the fermionic  operators can be combined into two Q-maps over $V_0$. 
So depending on the case, we obtain four to six
Q-maps over $V_0$ that generate the operator algebra in string theory, $\mathcal{A}$. These Q-maps can be combined into one Q-map, $\Phi_{7,k}$, 
over the test-function space  $V_0^{\otimes k}$ ($k = 4$ or $k=6$), i.e. $P(\Phi_{7,k})$ is the operator algebra in the respective string theory,
and we can apply theorem 1.\\[.3cm]
{\bf Corollary 9:} $\Phi_0$ is a core for $\Phi_{7,k}$.\\[.15cm]

\end{document}